\documentclass[11pt]{article}
\usepackage{graphicx}
\usepackage{epsfig}

\newcommand{\BABARPubYear}    {03}

\newcommand{\BABARConfNumber} {020}
\newcommand{\SLACPubNumber} {10107}

\providecommand{\Dspipi}{\ensuremath{D^{*+}\pi^-\pi^-}\xspace}

\input pubboard/babarsym

\usepackage{amsmath}

\def\Dm     {\ensuremath{D^-}\xspace}

\def\Dstarm {\ensuremath{D^{*-}}\xspace}

\providecommand{\Kpi}{\ensuremath{K^-\pi^+}\xspace}
\providecommand{\Kpipiz}{\ensuremath{K^-\pi^+\pi^0}\xspace}
\providecommand{\Kpipipi}{\ensuremath{K^-\pi^+\pi^-\pi^+}\xspace}
\providecommand{\Kspipi}{\ensuremath{\KS\pi^+\pi^-}\xspace}

\providecommand{\Kpipi}{\ensuremath{K^-\pi^+\pi^+}\xspace}
\providecommand{\Kspi}{\ensuremath{\KS\pi^+}\xspace}

\providecommand{\Dpipi}{\ensuremath{D^+\pi^-\pi^-}\xspace}
\providecommand{\Dspipi}{\ensuremath{D^{*+}\pi^-\pi^-}\xspace}

\providecommand{\de}{\ensuremath{\Delta E}\xspace}
\providecommand{\mes}{\ensuremath{m_{ES}}\xspace}

\providecommand{\Br}{\ensuremath{\mathcal{B}}\xspace}

\long\def\inst#1{\par\nobreak\kern 4pt\nobreak
    {\it #1}\par\vskip 10pt plus 3pt minus 3pt}

\begin{document}
{\pagestyle{empty}

\begin{flushright}
\babar-CONF-\BABARPubYear/\BABARConfNumber \\
SLAC-PUB-\SLACPubNumber \\
August 2003 \\
  \ \ \    \\
\end{flushright}

\par\vskip 5cm

\begin{center}
\Large {\bf Study of the Decays 
    $B^-\to D^{(*)+}\pi^-\pi^-$} \\
\end{center}
\bigskip

\begin{center}
\large The \babar\ Collaboration\\
\mbox{ }\\
\today
\end{center}
\bigskip \bigskip

\begin{center}
\large \bf Abstract
\end{center}

We report on analyses of $B^-$ mesons decaying into $D^{*+}\pi^-\pi^-$
and $D^+\pi^-\pi^-$ final states using 89 million $B^-$ decays 
collected by the \babar\ detector at the PEP-II asymmetric-energy $B$ Factory. 
Preliminary measurements are given for the inclusive branching fractions
for $B^- \to D^{*+}\pi^-\pi^-$ and $B^- \to D^+\pi^-\pi^-$, and for 
the exclusive branching fractions 
for $B^- \to D_1(2420)^0\pim$ and $B^- \to D_2^{*}(2460)^0\pim$, where 
$D_1(2420)^0$ and $D_2^{*}(2460)^0$ are the two narrow $c\bar u$ $P$-wave 
states. The ratio $\Br(\Bm \to D_2^*(2460)^0\pim)/\Br(\Bm \to 
D_1(2420)^0\pim)$ is measured to be $0.80 \pm 0.07 \pm 0.16$.

\vfill
\begin{center}
Contributed to the 
XXI$^{\rm st}$ International Symposium on Lepton and Photon Interactions at High~Energies, 8/11 --- 8/16/2003, Fermilab, Illinois USA
\end{center}

\vspace{1.0cm}
\begin{center}
{\em Stanford Linear Accelerator Center, Stanford University, 
Stanford, CA 94309} \\ \vspace{0.1cm}\hrule\vspace{0.1cm}
Work supported in part by Department of Energy contract DE-AC03-76SF00515.
\end{center}

\newpage
} 

\begin{center}
\small

The \babar\ Collaboration,
\bigskip

%
B.~Aubert,
R.~Barate,
D.~Boutigny,
J.-M.~Gaillard,
A.~Hicheur,
Y.~Karyotakis,
J.~P.~Lees,
P.~Robbe,
V.~Tisserand,
A.~Zghiche
\inst{Laboratoire de Physique des Particules, F-74941 Annecy-le-Vieux, France }
% ...

\end{center}\newpage

\section{INTRODUCTION}
\label{sec:Introduction}

Heavy Quark Effective Theory (HQET) takes as its starting point the
Heavy Quark Symmetry (HQS) limit in which the masses of both the initial- and
final-state heavy quarks in a decay are taken to be infinite
\cite{hqshqet,barteltshukla}.
Orbitally excited states of the $D$ meson, typically denoted $D_J$ (or
$D^{**}$), provide a unique opportunity to test HQET. 
The simplest $D_J$ meson consists of a charm quark and a 
light quark in an orbital angular momentum $L=1$ ($P$-wave) state.
The spectroscopy of $P$-wave $D$ mesons is summarized in 
Ref.~\cite{barteltshukla}. In the HQS limit ($m_c \gg \Lambda_{QCD}$), 
analogous to the hydrogen atom, the spin of the 
charm quark decouples from the other angular momenta, the angular 
momentum sum ${\bf j}= {\bf S}_q + {\bf L}$ of the light quark spin 
${\bf S}_q$ and the orbital angular momentum ${\bf L}$ is conserved, and 
$j$ is a good quantum number. Therefore, one expects one doublet of states 
with $j=3/2$ and one doublet with $j=1/2$~\cite{infin_mass}.
In the limit of a light charm quark ($m_c \ll \Lambda_{QCD}$), analogous to 
positronium, the $L=1$ combines with the total spin of $S=0$ or 1 to produce 
a singlet state with $J=1$ and $S=0$, and a triplet of states 
with $J=0,1,2$ and 
$S=1$~\cite{barteltshukla}. The $J=0$ state must have $j=1/2$ and the $J=2$ 
state $j=3/2$. However, the two $J=1$ states may be a mixture of $j=1/2$ and 
$j=3/2$. The details of this mixing probe the breaking of the Heavy Quark 
Symmetry due to the finite mass of the charm quark~\cite{isgurwise89}.

\begin{figure}[h]
\centering
\epsfig{width=0.70\textwidth,file=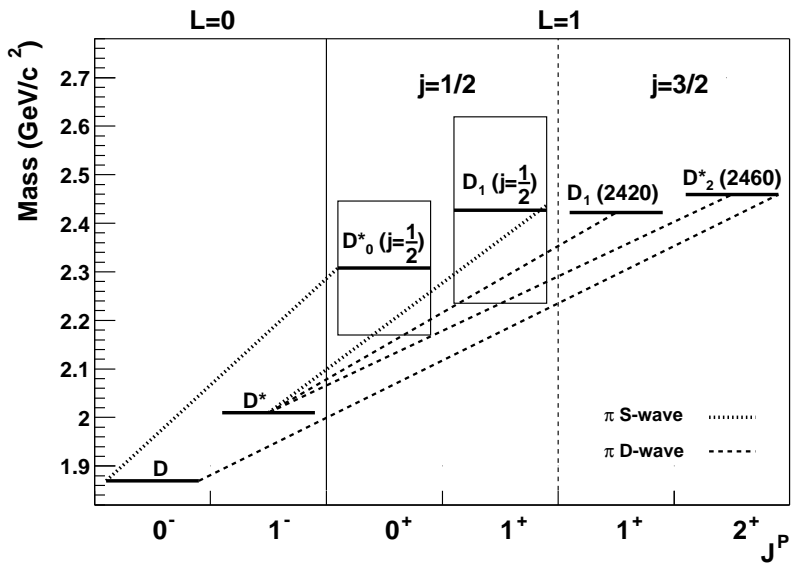}
\caption{Mass spectrum for $c\bar{q}$ states. The open boxes
 indicate that the $D_0^*(j=1/2)$ and $D_1(j=1/2)$ 
 are expected to be wide. Lines between levels show anticipated pion
 transitions. Narrow masses are from Ref.~\cite{PDG}, wide masses and widths
 are from Ref.~\cite{belle-prd}.
  } 
\label{fig:spec}
\end{figure}

We label the four neutral $D$-meson $P$-wave 
states\footnote{Charged conjugate states 
are implied throughout the paper.} as $D_0^*(j=1/2)^0$, 
$D_1(2420)^0$, $D_1(j=1/2)^0$, and $D_2^*(2460)^0$ 
and show the mass spectrum and expected
transitions in Fig.~\ref{fig:spec}. 
The conservation of parity and angular momentum restricts the final
states and partial waves that are allowed in the decays of the various
$D_J$ mesons. The resonances that decay through a $D$-wave are expected to
be narrow (20--30 \mev) and the resonances that decay through an
$S$-wave are expected to be wide (a few hundred \mev). 
The $D^*_2(2460)^0$ can
only decay via a $D$-wave and the $D^*_0(j=1/2)^0$ can only decay via an
$S$-wave. The $D_1(j=1/2)^0$ and $D_1(2420)^0$ may decay via $S$-wave or
$D$-wave. In the HQS limit, 
the $D_1(2420)^0$ will decay only via $D$-wave and is therefore a narrow 
resonance. Analogously, the $D_1(j=1/2)^0$ will decay only
via $S$-wave and is therefore a broad resonance.

The members of the isospin doublet of narrow resonances, $D^*_2(2460)^0$
and $D_1(2420)^0$, have been observed by many experiments~\cite{prevmeas}.
The properties of the $D_J^0$ mesons as listed in the 2002 Review of
Particle Physics are given in Table~\ref{tab:dstst-properties}.

\begin{table}[htbp]
  \begin{center}
  \caption{Properties of $L=1$ $D_J^0$ mesons. Masses and widths are from 
    Ref.~\cite{PDG}.\label{tab:dstst-properties}}
  \vspace{0.1in}
  \begin{tabular}{lcccccc}
    \hline
    \multicolumn{1}{c}{State} & $J^P$ & Mass & Width & Decays & Partial & HQS \\
                      &   & (\mevcc) & (\mev) & & waves & allowed \\
    \hline
  $D^*_0(j=1/2)^0$ & $0^+$ & --- & --- & $D\pi$ & S & S \\ 
  $D_1(2420)^0$    & $1^+$ & $2422.2\pm 1.8$& $18.9^{+4.6}_{-3.5}$ & $D^*\pi$ & S,D & D \\ 
    $D_1(j=1/2)^0$ & $1^+$ & --- & --- & $D^*\pi$ & S,D & S \\ 
    $D^*_2(2460)^0$ & $2^+$ & $2458.9\pm 2.0$& $23\pm 5$ & $D^*\pi,\ D\pi$ & D & D \\ 
    \hline
  \end{tabular}
  \end{center}
\end{table}

Predictions based on HQET can differ due to variations in the calculational
techniques employed. One such case involves the ratio 
of the branching fractions, \Br, for the two narrow states
\begin{equation}
R \equiv {{\Br(\Bm \to D_2^*(2460)^0\pim})\over{\Br(\Bm \to D_1(2420)^0\pim)}}.
\label{eq:R}
\end{equation}
Reference~\cite{leibovich}
obtains values of $R$ between 0 and 1.5.
Reference~\cite{neubert} predicts $R \approx 0.35$.

In this paper we present preliminary measurements of the inclusive branching 
fractions for $B^- \to D^{*+}\pi^-\pi^-$ and $B^- \to D^+\pi^-\pi^-$,
the exclusive branching fractions for $B^- \to D_1(2420)^0\pim$ and 
$B^- \to D_2^{*}(2460)^0\pim$, and $R$.

\section{THE \babar\ DETECTOR AND DATASET}
\label{sec:babar}

The data used in this analysis were collected with the \babar\ detector
at the \pep2\ $\epem$ asymmetric-energy storage ring during the 
years 1999 - 2002.  
The sample corresponds to an integrated luminosity of $81.9~\invfb$ 
accumulated on the \FourS\ resonance (``on-resonance'') and 
$9.6~\invfb$ accumulated at an $e^+e^-$ 
center-of-mass (CM) energy about $40\mev$ below the \FourS\ resonance
(``off-resonance''), which are used for $q\bar q$ background studies.
The on-resonance sample corresponds to $(88.9 \pm 0.9)$ million
\BB\ pairs.

A detailed description of the \babar\ detector can be found elsewhere  
~\cite{babarnim}. Charged particle trajectories are measured by a five-layer
double-sided silicon vertex tracker (SVT) and a 40-layer drift chamber (DCH), 
which lie within a 1.5~T solenoidal magnetic field. Charged particle 
identification is achieved by combining ionization-energy loss ($dE/dx$) 
measurements in the DCH and SVT with information from the ring-imaging
Cherenkov detector. Photons are identified in a CsI(Tl) electromagnetic
calorimeter. The instrumented flux return is equipped with resistive plate
chambers for muon and neutral hadron identification.

\section{ANALYSIS METHOD}
\label{sec:Analysis}

We reconstruct the $B^-$ decays to the final states 
$D^{*+} \pi^-\pi^-$ and $D^{+} \pi^-\pi^-$ for study of the properties of 
the $D_J^0$ resonances. $D^{*+}$ candidates are reconstructed by combining 
$D^0$ candidates and a \pip. $D^0$ candidates are reconstructed in 
the modes: $D^0 \to K^- \pi^+$, $D^0 \to K^- \pi^+ \pi^0$, 
$D^0 \to K^- \pi^+ \pi^- \pi^+$, and $D^0 \to K_s^0 \pi^+ \pi^-$.
$D^+$ candidates are reconstructed in the modes: 
$D^+ \to K^- \pi^+ \pi^+$ and $D^+ \to K_s^0 \pi^+$.
As indicated in Table~\ref{tab:dstst-properties}, we expect the decay 
$\Bm\to D^{*+} \pi^-\pi^-$ to proceed dominantly through the intermediate 
states $D^*_2(2460)^0\pi^-$, $D_1(2420)^0\pi^-$, and $D_1(j=1/2)^0\pi^-$, 
which involve the two narrow $D_J^0$ resonances and the wide 
$D_1(j=1/2)^0$ resonance respectively. A contribution
from the three-body decay $B^-\to D^{*+} \pi^-\pi^-$ (nonresonant) is 
also possible.
The decay $B^-\to D^{+} \pi^-\pi^-$ is expected to proceed primarily through
the intermediate states $D^*_2(2460)^0\pi^-$ and $D_0^*(j=1/2)^0\pi^-$ with a 
contribution from three-body nonresonant decay of $\Bm\to D^{+} \pi^-\pi^-$.

The selection criteria are optimized by maximizing $S^2/(S+B)$, where the 
signal yield $S$ is based on signal Monte Carlo and the branching 
fractions presented by the BELLE collaboration~\cite{belle-conf-0235}, and 
the expected background yield $B$ is determined from generic $B\bar{B}$ MC 
(after removing the signal events) and off-resonance data.

\begin{table}
\begin{center}
\caption{Summary of selection criteria for each mode. Masses are given 
in MeV/$c^2$, energies are in MeV, and the $K_s^0$ flight length is in cm.
The $|\Delta E |$ requirement is not applied for the measurements of the 
inclusive 
$B^-\to D^{*+}\pi^-\pi^-$ and $B^-\to D^{+}\pi^-\pi^-$ branching fractions.
The selection criteria are described in more detail in the text.
\label{table:optimize}}
\vspace{0.1in}
\begin{tabular}{l|cccc|cc}
\hline
 & \multicolumn{4}{c|}{$B\to\Dspipi$, $D^{*+} \to D^0\pip$} 
  & \multicolumn{2}{c}{$B\to\Dpipi$}\\

 & \multicolumn{4}{c|}{$D^0 \to$} 
  & \multicolumn{2}{c}{$D^+ \to$}\\

Selection Criteria & \Kpi & \Kpipiz & \Kpipipi & \Kspipi & \Kpipi &\Kspi\\
\hline
$|\cos \theta_{\rm thrust}|$ & $<0.95$ & $<0.95$ & $<0.95$ & $<0.95$ 
        & $<0.95$ & $<0.95$\\
$R_2$ & $<0.8$ & $<0.7$ & $<0.8$ & $<0.7$& $<0.35$ & $<0.35$\\
$|\delta m(\piz)|$   & --- & $<18$ & --- & --- & --- & ---\\
$E_{\rm min}(\gamma)$ & --- & $>35$ & --- & --- & --- & ---\\
$|\delta m(\KS)|$               & --- & --- & --- & $<9$ & --- & $<9$\\
\KS flight length      & --- & --- & --- & $>0.1$ & --- & $>0.1$\\
Dalitz weight        & --- & $>0.01$ & --- & --- & --- & ---\\
$|\delta[m(D^{*+})-m(D^0)]|$ & $<3.0$ & $<2.6$ & $<2.8$ & $<3.0$ & --- & ---\\
$|\delta m(D^{0})|$  & $<31$  & $<29$  & $<15$ & $<21$ & --- & ---\\
$|\delta m(D^{+})|$ & --- & --- & --- & --- & $<15$ & $<15$\\
\mes & $>5274$ & $>5274$ & $>5274$ & $>5274$ & $>5274$ & $>5271$\\
\hline
$|\de|$ & $<28$& $<26$ & $<24$ & $<32$ & $<26$  & $< 28$\\
\hline
\end{tabular}
\end{center}
\end{table}

The complete set of selection criteria is given in Table~\ref{table:optimize}. 
Suppression of background from nonresonant $q\bar q$
production is provided by two topological requirements.
In particular, we employ restrictions on the magnitude of the cosine of the
thrust angle, $\theta_{\rm thrust}$, defined as the angle between the thrust 
axis of the particles that form the reconstructed $B^-$ candidate and the 
thrust axis of the remaining tracks and neutral clusters in the event.
We also select on the ratio of the second to the zeroth Fox-Wolfram 
moment~\cite{foxwolfram},
$R_2$, to gain additional discrimination between signal events and those from 
continuum background. 
Neutral pion candidates are used in the reconstruction of $D^0$ via the 
$K \pi \piz$ decay mode and are selected based on the \piz mass,
$m(\piz)$, and the minimum
photon energy, $E_{\rm min}(\gamma)$.  
Kaon track selection is based on information from the tracking and DIRC
systems and has an efficiency of about $80\%$.
Charged pion tracks in all modes are required {\bf not} to satisfy the 
kaon track selection.
The \KS candidates are reconstructed using two charged tracks with an 
invariant mass, $m(\KS)$, within a region around the
nominal \KS mass. In addition, a constraint on the flight distance of 
the \KS candidate from the primary vertex significantly reduces background
from combinatorics.
For the $D^0 \to K^- \pi^+ \pi^0$ mode, intermediate resonant states, such as
$K^-\rho^+$, 
$K^{*0}$(892)$\pi^0$ and $K^{*-}$(892)$\pi^+$, are exploited 
by the use of a Dalitz weight criteria which selects 71\% of real
$D^0 \to K^- \pi^+ \pi^0$ events and rejects 72\% of fake 
$D^0 \to K^- \pi^+ \pi^0$ events.
The masses of the $D^0$ and $D^+$ candidates are required to be in a region 
around their nominal values. For $D^{*+}$ candidates, the mass difference, 
$\Delta m = m(D^{*+}) - m(D^0)$, is required to be in a region around 
the nominal value. The widths of these regions are dependent on the decay 
mode of the $D^0$ or $D^+$ and are determined from the optimization 
(see Table~\ref{table:optimize}).

\Bm candidates are reconstructed by combining either a $D^{*+}$ or $D^+$ 
candidate with two \pim mesons. The standard kinematic variables, \mes and 
$\DeltaE$, define the signal region and are computed as 
follows:
\begin{eqnarray}
m_{\rm ES}&=& \sqrt{(E_{\rm beam}^*)^2 - (\sum_i {\bf p}^*_i)^2}, \\
\Delta E&=& \sum_i\sqrt{m_i^2 + ({\bf p}_i^*)^2}- E_{\rm beam}^*,
\end{eqnarray}
where $E_{\rm beam}^*$ is the beam energy in the \FourS\ CM frame,
${\bf p}_i^*$ is the CM momentum of particle $i$ in the
candidate $B^-$-meson system, and $m_i$ is the mass of particle $i$.
For signal events, the beam-energy-substituted $B^-$ mass, $m_{\rm ES}$, 
peaks at $m_B$. The quantity $\Delta E$ is used to determine
whether a candidate system of particles has total energy consistent with
the beam energy in the CM frame.

\begin{figure}[t]
\centering
\epsfig{width=0.49\textwidth,file=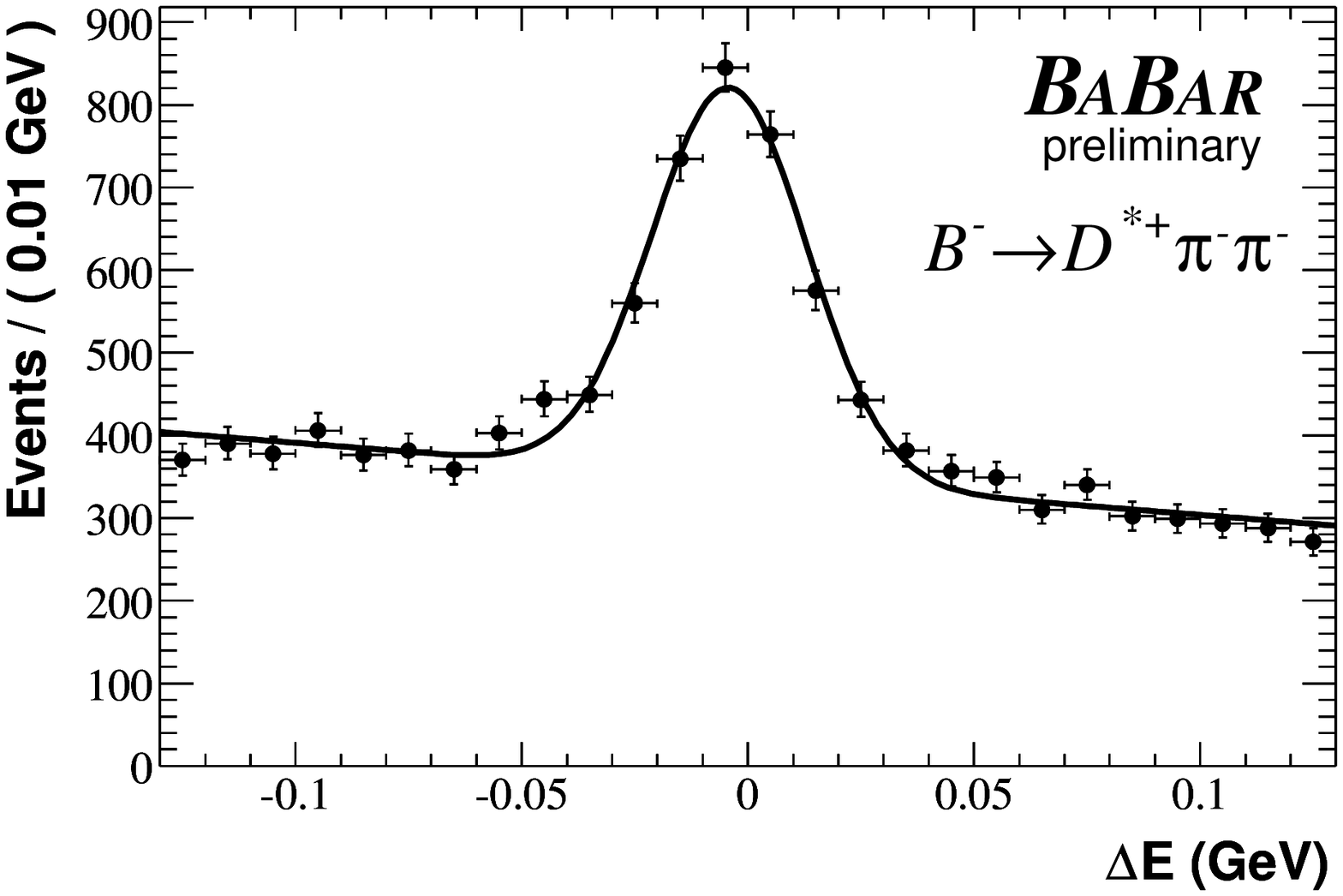}
\epsfig{width=0.49\textwidth,file=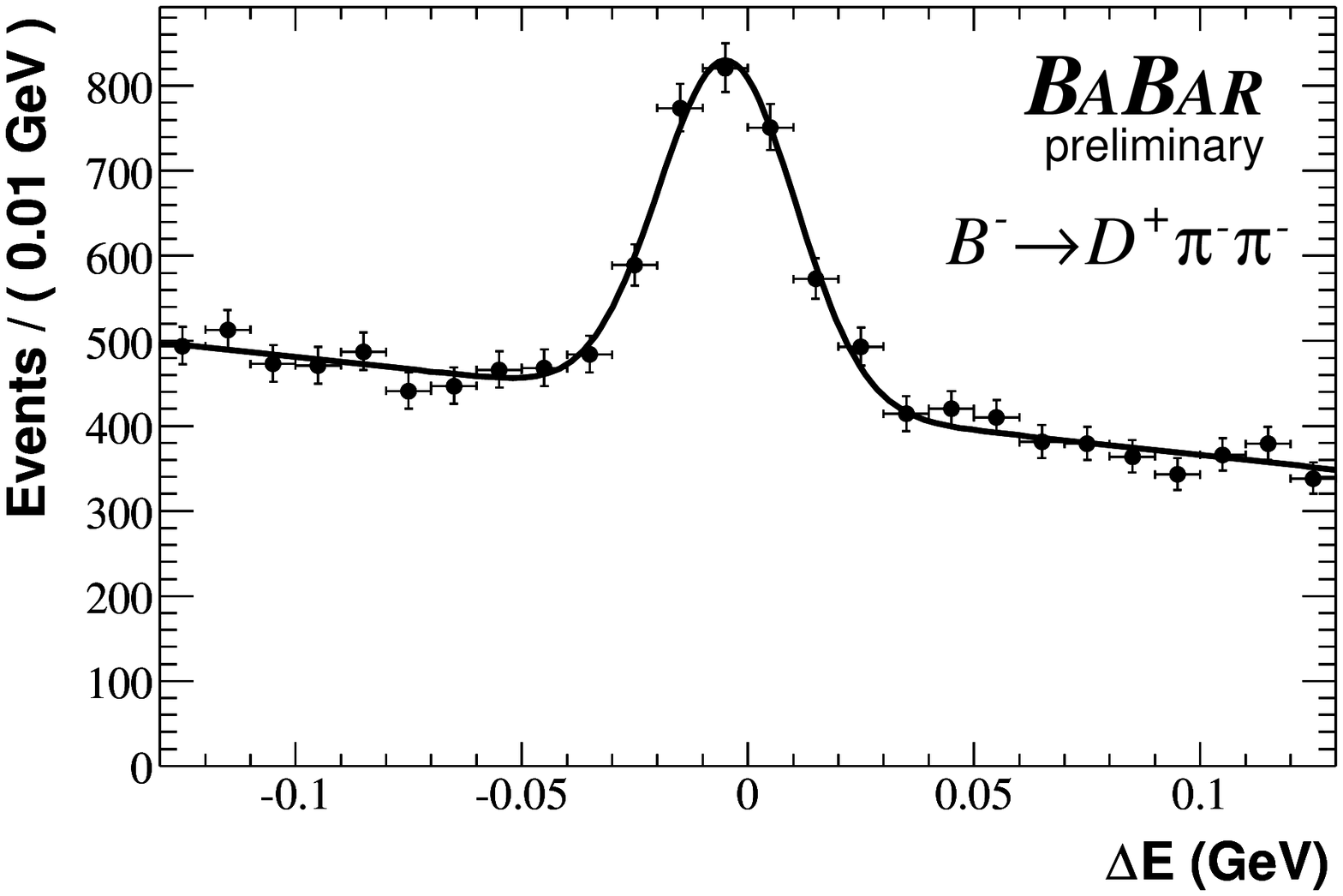}
\caption{Distributions of $\Delta E$ for the selected $\Bm \to
  D^{*+} \pim \pim$ (left) and $\Bm \to D^+ \pim \pim$ (right)
  candidates in the \mes signal region. The signal is fit with a 
  Gaussian and the background is fit
  with a linear function.}  
\label{fig:defit}
\end{figure}

For each decay mode, we fit the \de distributions of the selected 
$\Bm\to D^{*+}\pim\pim$ and $\Bm\to D^{+}\pim\pim$ candidates in the 
\mes signal region  (see Table~\ref{table:optimize})
with a Gaussian for the signal and a linear function for the background.
The resolution of the \de distributions is about 18 MeV.
The number of events with multiple $B$ candidates 
is found to be about $15\%$ for the mode containing
$\pi^0$ and $2-5\%$ for the rest.
The results of the fit are shown in Table~\ref{table:dEfit} and 
Fig.~\ref{fig:defit}.

The inclusive $\Bm \to D^{(*)+} \pim \pim$ branching fraction for each 
$D^0/D^+$
decay mode $k$ ($k=\Kpi,$ $\Kpipiz$, $\Kpipipi$, and $\Kspipi$ for $D^0$,
and $k = \Kpipi,$ $\Kspi$ for $D^+$) is given by the relation 
\begin{equation}
\Br_k = {N_{{\rm signal},k}\over{(\epsilon_k \cdot 
\Br(D^{(*)+})_k)\cdot N(B^-)}},
\label{eq:brratio}
\end{equation}
where $N_{{\rm signal},k}$ is the fitted signal yield,
$\epsilon_k$ is the $\Bm \to D^{(*)+}\pim\pim$ reconstruction 
efficiency, as determined from Monte Carlo simulations, averaged over 
all contributing 
$D_J^0$ resonances and nonresonant three-body decay, 
$\Br(D^{*+})_k$ is the product of the branching ratio for 
$D^{*+}\to D^0\pip$ $(=(67.7\pm0.5)\%)$ 
and the $D^0$ branching fraction for decay mode $k$, $\Br(D^+)_k$ is the 
$D^+$ branching fraction for decay mode $k$~\cite{PDG},
$N(B^-) = (88.9 \pm 0.9) \times 10^6$. 
The values of these quantities, along with the 
resultant branching fraction for each decay mode, are given in 
Table~\ref{table:dEfit}. The average 
$\Bm \to D^{*+}\pim\pim$ and $\Bm \to D^{+}\pim\pim$ 
branching ratios, which are calculated as the
weighted mean of the branching ratios measured for the different $D^0$ or 
$D^+$ decay modes, are also given. The weight for each $D^0$ or $D^+$ 
decay mode is the inverse of the statistical uncertainty for that mode.

\begin{table}[h]
\begin{center}
\caption{Number of signal candidates from the fit to the $\Delta E$ 
 distribution, average efficiency, and measured branching fraction. 
 Uncertainties are statistical only.
\label{table:dEfit}}
\vspace{0.1in}
\begin{tabular}{l|cc|c}
\hline
Decay Mode & $N_{\rm Signal}$& 
Avg. $\epsilon_k$ (\%) & Br. Frac. (\%)\\
\hline
\hline
\multicolumn{4}{c}{$\Bm \to D^{*+}\pim\pim$}\\
\hline
\hline
$D^0 \to$ all modes & 
           $  1997 \pm 81$ & -- & $ 1.22 \pm 0.05 $ \\
\hline
$D^0 \to K^- \pi^+$ & 
 $   571 \pm 36$ & $21.3 \pm 0.9$ & $   1.17 \pm 0.07 $ \\
$D^0 \to K^- \pi^+ \pi^0$ & 
 $   735 \pm 56$ & $7.6 \pm 0.4$ & $   1.23 \pm 0.09 $ \\
$D^0 \to K^- \pi^+ \pi^- \pi^+$ & 
 $   627 \pm 42$ & $9.9 \pm 0.5$ & $   1.41 \pm 0.09 $ \\
$D^0 \to K_s^0 \pi^+ \pi^-$ &
 $    48 \pm 13$ & $5.5 \pm 0.5$ & $   0.72 \pm 0.19 $ \\

\hline
\hline
\multicolumn{4}{c}{$\Bm \to D^+\pim\pim$}\\
\hline
\hline
$D^+ \to$ all modes & 
          $  1514 \pm 78$ & -- & $   0.87 \pm 0.04 $ \\
\hline
$D^+ \to K^- \pi^+ \pi^+$ & 
 $  1417 \pm 76$ & $20.4 \pm 0.6$ & $   0.86 \pm 0.05 $ \\
$D^+ \to K_s^0 \pi^+$ & 
 $    88 \pm 17$ & $10.5 \pm 0.5$ & $   1.00 \pm 0.19 $ \\
\hline
\end{tabular}
\end{center}
\end{table}%

\begin{figure}[h]
\centering
\epsfig{width=0.49\textwidth,file=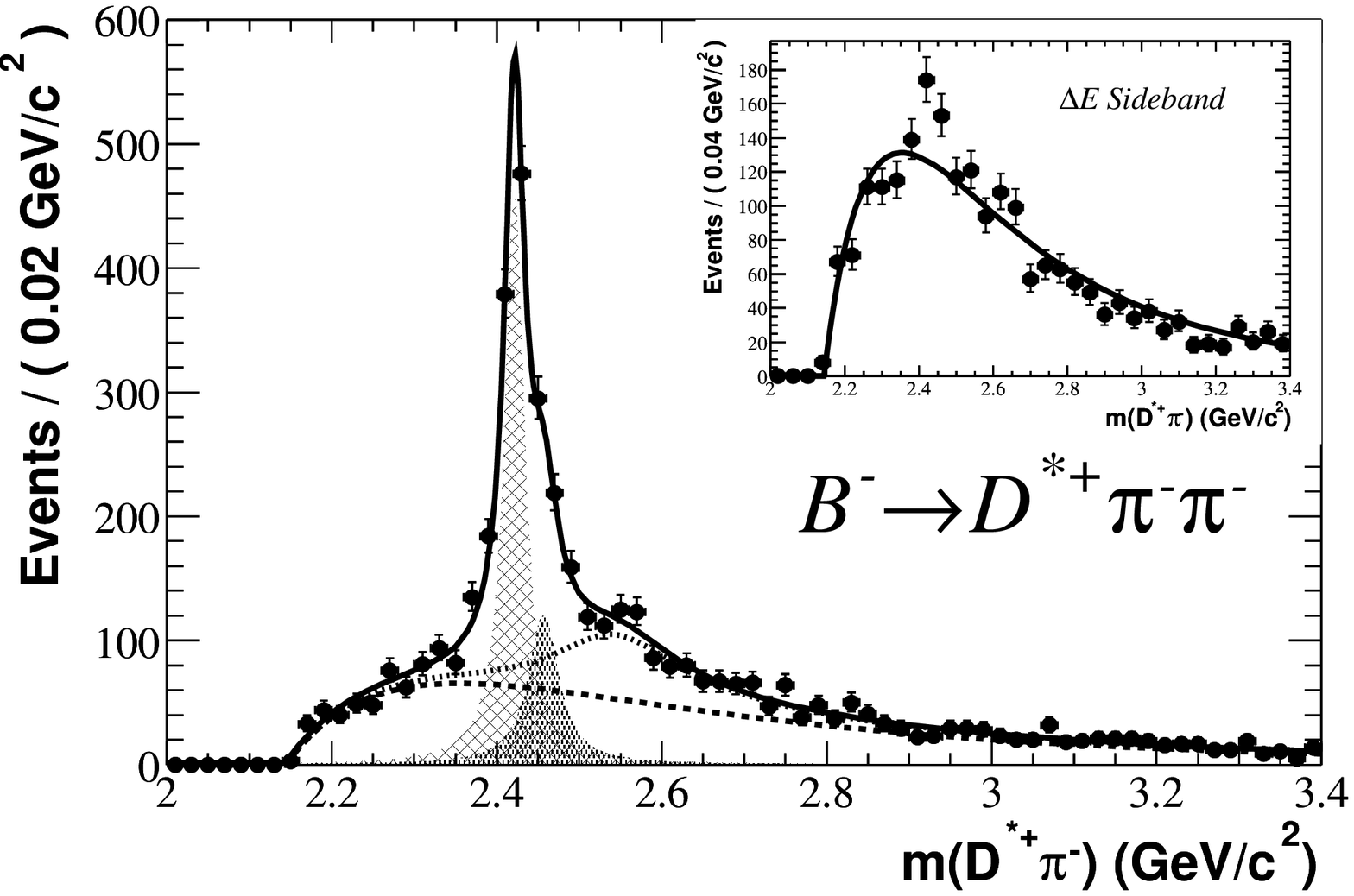}
\epsfig{width=0.49\textwidth,file=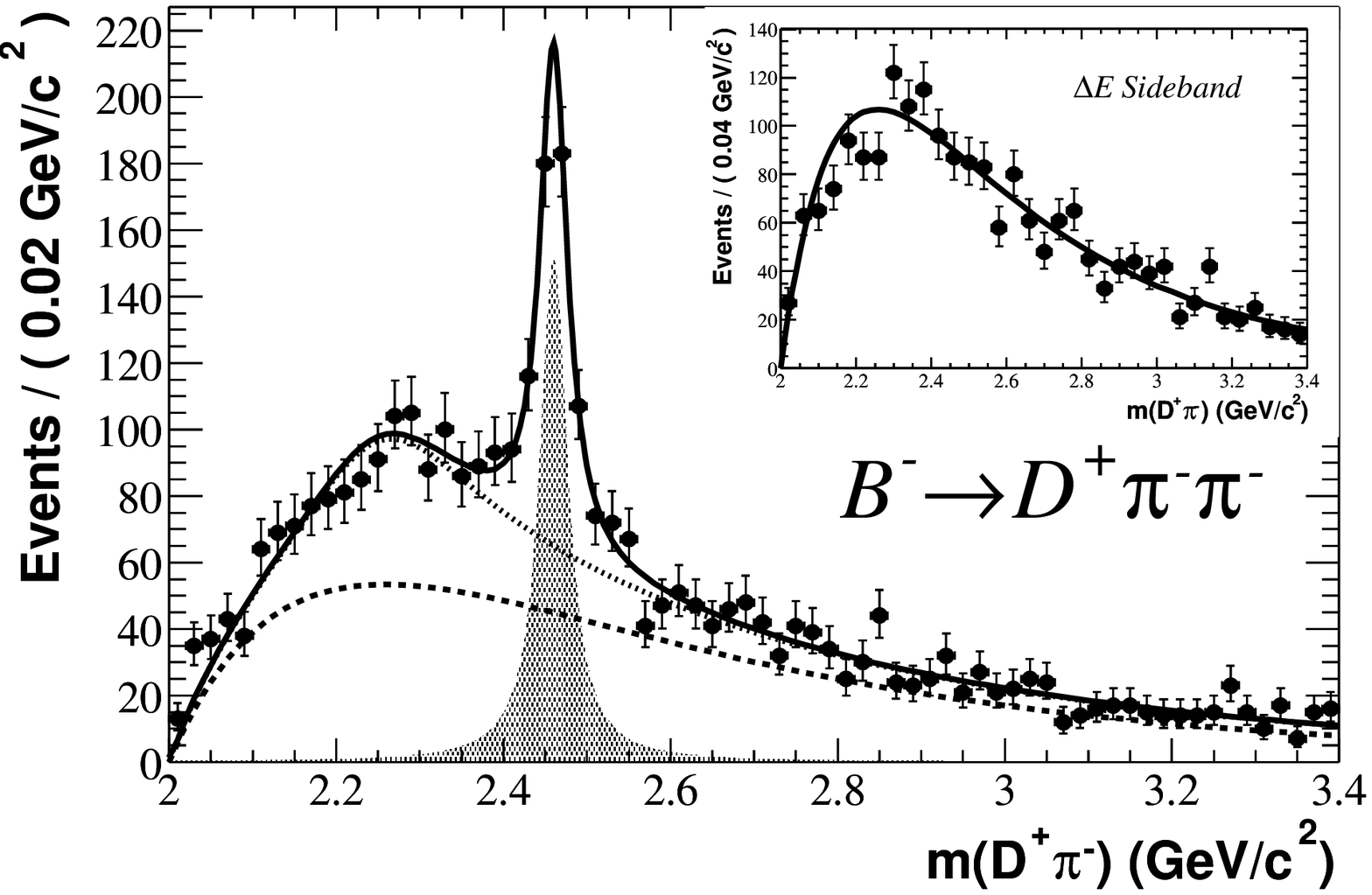}
\caption{
$m(D^{(*)+}\pim)$ shown for the $D^{*+} \pim \pim$ candidates (left) and 
$D^+ \pim \pim$ candidates (right). 
The data corresponding to the signal region is shown overlaid 
with the fit result. 
A fit to the data in the $\Delta E $ sideband region is shown with a dashed 
line. 
An enlarged view of the background shape can be seen in the inset, 
which shows the $\Delta E $ sideband region with a larger 
($\times 2$) bin size.
The dotted line shows a fit to the sum of the background (dashed line) and 
the contribution from the wide $D_J$ states.
The contributions from the $D_1(2420)^0$ and $D^*_2(2460)^0$ are shown 
with hashed and solid areas, respectively. The fitting procedure is
described in the text.
  } 
\label{fig:d2fit}
\end{figure}

For each $\Bm$ candidate in the (\mes, \de) signal region (as defined in 
Table~\ref{table:optimize}), the exclusive $D_J$ candidates are formed,  
by combining the $D^{(*)+}$ daughter with the slower of the two 
pions from the $B^-$ decay.
The $D^{*+}\pim$ and $D^{+}\pim$ mass distributions are shown 
for the $\Bm \to \Dspipi$ and $\Bm \to \Dpipi$ final states respectively 
in Figure~\ref{fig:d2fit} summed over all $D^0/D^+$ decay modes.
In this preliminary analysis, 
we do not attempt to fit the $\Bm \to D^{(*)+}\pim\pim$ Dalitz distributions.
Upper and lower \de sidebands, defined by $|\de| > 60$~MeV, each with 
half the width of signal region, are used to estimate the 
background under the $D_J$ mass peak.

Assuming the decay chain
$\Bm \to D_J^0\pim \to D^{(*)+}\pim\pim$, we perform an unbinned 
maximum likelihood fit to the $m(D^{(*)+}\pim)$ distribution of the 
$\Delta E$ signal region and sidebands to obtain the mass and 
width of each narrow $D_J^0$ resonance. In the fit we
describe the signal $m(D^{(*)+}\pim)$ distribution with a 
non-relativistic Breit-Wigner
shape convolved with a Gaussian function that represents the detector
resolution. 
For the $\Bm \to D^{*+}\pim\pim$ case, we fit simultaneously all 
$D^0$ decay modes to extract the mass and width of both the 
$D_1(2420)^0$ and the $D^*_2(2460)^0$.
For the $\Bm \to D^{+}\pim\pim$ case, we fit simultaneously all $D^+$ 
decay modes to extract the mass and width of 
the $D^*_2(2460)^0$.
For the present analysis, the broad $D_J^0$ resonances and 
nonresonant \Bm decay modes are not distinguished from each other. 
In the fit for each decay mode, the broad resonances are described by a 
relativistic Breit-Wigner shape. 
The combinatorial background, determined primarily by events in 
the $\Delta E$ sidebands, is described by a threshold function. 
The small excess at the signal mass seen in the inset plot 
for the $D^{*+}\pim\pim$ mode in Fig.~\ref{fig:d2fit}, showing data from the 
\de sideband region, is found from Monte Carlo studies to be due to self 
crossfeed from true $D_J$ states that are 
combined with a wrong pion and which therefore misreconstruct the $B^-$.
We account for the excess as a systematic uncertainty.

To obtain the yield for each $D^0/D^+$ decay mode,
similar fits are performed on each mode 
separately with the mass and width values of the narrow resonances
($D_1(2420)^0$ and $D^*_2(2460)^0$) fixed to those found from the 
global fit over all modes. Since the mass and width of the $D^*_2(2460)^0$
are found from both the $m(D^{*+}\pim)$ and $m(D^{+}\pim)$ global fits, the
fixed values used are the weighted average of the two.

For each $D^0$ or $D^+$ decay mode, the calculation of the exclusive branching 
ratios from the yields and efficiencies is analogous to the 
inclusive case (see Eq.~\ref{eq:brratio}). The exclusive efficiencies,
as determined from Monte Carlo simulations, and the yields and resultant 
branching fractions for each decay mode are given in 
Table~\ref{table:d2fitSignal}. The weighted averages of the branching ratios 
found for the individual $D^0$ and $D^+$ decay modes are also given.

\begin{table}[h]
\begin{center}
\caption{Reconstruction efficiencies for $m(D^{*+}\pim)$ and 
$m(D^{+}\pim)$ fits for each mode as determined from $\Bm\to D_J^0\pi^-$ 
simulated events, $D_J^0$ signal yields, and $D_J^0$ branching ratios for 
each resonance. ``$D^0/D^+ \to {\rm all \ modes}$'' branching fractions 
correspond to a weighted average of the individual modes.
\label{table:d2fitSignal}}
\begin{tabular}{l|cc|cc|cc}
\hline
 & \multicolumn{2}{c}{Reconstruction Eff.} 
 & \multicolumn{2}{|c}{Yield} 
 & \multicolumn{2}{|c}{Br. Ratio ($\times 10^{-3}$)}\\
Decay Mode & $D_1(2420)^0$ & $D_2^*(2460)^0$ & 
$D_1(2420)^0$ & $D_2^*(2460)^0$ &
$D_1(2420)^0$ & $D_2^*(2460)^0$ \\
\hline
\hline
 & \multicolumn{6}{c}{$\Bm\to D_J^0 \pi^- \to D^{*+}\pim\pim$}\\
\hline
$D^0 \to$ all modes             
&               &
& $ 887 \pm 44$ & $  307 \pm 41$ 
& $0.59 \pm 0.03$ & $0.18 \pm 0.03$ \\
\hline
$D^0 \to K^- \pi^+$             
& $19.8 \pm 0.3$  & $20.9 \pm 0.4$
& $ 227 \pm 22$ & $  110 \pm 21$ 
& $0.50 \pm 0.05$ & $0.23 \pm 0.04$ \\
$D^0 \to K^- \pi^+ \pi^0$       
& $7.0 \pm 0.2$ & $7.2 \pm 0.2$
& $ 319 \pm 27$ & $   94 \pm 25$ 
& $0.58 \pm 0.05$ & $0.17 \pm 0.04$ \\
$D^0 \to K^- \pi^+ \pi^- \pi^+$ 
& $8.8 \pm 0.2$ & $9.4 \pm 0.2$
& $ 304 \pm 26$ & $   97 \pm 23$ 
& $0.77 \pm 0.07$ & $0.23 \pm 0.05$ \\
$D^0 \to K_s^0 \pi^+ \pi^-$     
& $5.9 \pm 0.2$ & $6.1 \pm 0.2$
& $  44 \pm  8$ & $    0 \pm  5$ 
& $0.61 \pm 0.11$ & $0.00 \pm 0.07$ \\
\hline
 & \multicolumn{6}{c}{$\Bm\to D_J^0 \pi^- \to D^{+}\pim\pim$}\\
\hline
$D^+ \to$ all modes       
&               &
& --- & $  465 \pm 33$ 
& --- & $0.29 \pm 0.02$ \\
\hline
$D^+ \to K^- \pi^+ \pi^+$ 
& ---           & $18.3 \pm 0.3$
& --- & $  430 \pm 32$ 
& --- & $ 0.29 \pm 0.02$ \\
$D^+ \to K_s^0 \pi^+$     
& ---           & $9.6 \pm 0.2$
& --- & $   30 \pm  8$ 
& --- & $0.37 \pm 0.10$ \\
\hline
\end{tabular}
\end{center}
\end{table}

Quantitative study
of the wide resonances, 
$D_0^*(j=1/2)^0$ and $D_1(j=1/2)^0$, requires a Dalitz analysis and will 
form the basis of a subsequent publication. Interference between 
the two $D_1^0$ resonances in the decay $\Bm \to D^{*+}\pim\pim$ 
is, therefore, not within the scope of the present paper.

\section{STUDIES OF SYSTEMATIC UNCERTAINTIES}
\label{sec:Systematics}

As listed in Table~\ref{tab:syserr}, the systematic uncertainties on the
measurement of the inclusive $\Bm \to D^{(*)+}\pim\pim$ and exclusive 
$\Bm \to D_J^0\pim$ branching fractions are due to sources such as
tracking and $\pi^0$ reconstruction efficiencies, particle 
identification, input branching fractions, and $B$-counting, and from 
uncertainties in modeling the wide resonances, 
our methods for determining the efficiencies for reconstructing resonant 
and nonresonant decays, and in fitting the narrow resonances.

Because very little is known of the wide
resonances, $D_0^*(j=1/2)^0$ and $D_1(j=1/2)^0$, and the nonresonant
decays of \Bm, describing the shape of the wide resonances
is difficult.  
To estimate the uncertainty due to our description, 
we fix the mass and the width of the wide resonances
to those of the BELLE Collaboration~\cite{belle-conf-0235},
and the differences between fitting with fixed mass and width of the wide 
resonance and fitting with floating mass and width are taken as a systematic
uncertainty.
We estimate a $4\%$ systematic uncertainty due the combination
true $D_J$ states with the incorrect pion.
We estimate a contribution to the systematic uncertainty due to multiple
signal candidates in an event from the difference in the branching
fractions between using the candidate with $\mes$ closest to the nominal
value and using all candidates. 
We estimate the uncertainty in our determination of the efficiencies of 
the resonant and nonresonant states from the spread among the efficiencies,
for each decay mode, of the various resonant and nonresonant states 
(as determined from Monte Carlo simulations). 
To estimate the uncertainty on the yields from our fits to the narrow 
resonances, we vary the masses and widths of the $D_1(2420)^0$ and 
$D^*_2(2460)^0$ in the
fits to the $m(D^{(*)+}\pim)$ distributions in each $D^0$ and $D^+$
decay mode by one sigma around their values as determined in the global 
(all decay modes included) fits.
Uncertainties on the input branching fractions are taken from the 
Review of Particle Physics~\cite{PDG}.

\begin{table}[!htbp]
 \begin{center}
 \caption{Sources of systematic uncertainties in fractions. 
``Bachelor'' pions are those (two) pions that do not originate from
the decay of a $D^{*+}$, $D^+$, or $D^0$. 
The upper range of the uncertainties is dominated by 
contributions from the $K_s$ modes which have only a small contribution
to the total branching fraction measurements 
(see Table~\ref{table:d2fitSignal}).
For the $\Bm \to \Dp\pim\pim$, 
$\Dp \to K_s^0\pip$ mode for the $D_2^*(2460)^0$, this uncertainty is 23\%.
 \label{tab:syserr}}
 \vspace{0.1in}
 \begin{tabular}{l|c}
 \hline
 \multicolumn{2}{c}{Correlated Systematic Uncertainties} \\
 \hline
 Uncertainties on Tracking efficiency: & \\ 
      \hskip 1.0 cm   bachelor pions & $3.5\%$   \\ 
      \hskip 1.0 cm all other tracks & 1.3\%     \\
 Uncertainties on Particle ID (kaon efficiency)   & $2.5\%$ \\
 Efficiency difference among the 
 resonant and nonresonant decay: & \\
      \hskip 1.0 cm $K^-$ modes             & $(3 - 4)\%$   \\
      \hskip 1.0 cm $K_s$ modes           & $(8 - 13)\%$  \\
 Multiple $B$ candidates & $ (2 - 15)\%$ \\
 $B$-counting & $1.1\%$ \\
 $D^{*+}$ branching fractions for \Dspipi results & $0.7 \%$ \\
 \hline
 \multicolumn{2}{c}{Uncorrelated Systematic Uncertainties} \\
 \hline
 \piz efficiency & 7.7\% \\
 \KS efficiency &        $3\%$ \\
 $D^0$ and $D^+$ branching fractions & ($2.3 - 6.2)\%$ \\
 Monte Carlo statistics & ($1.5 - 5.2)\%$ \\
 $K^0\to\KS\to\pip\pim$ branching fraction & $0.4 \%$ \\
 \hline
 \multicolumn{2}{c}{Systematic Uncertainties on Exclusive BF } \\
 \hline
  Uncertainties in the description of the wide resonances &
  $(4.5 - 11.8)\%$ \\
  Peaking background from real $D_J$ + wrong $\pi$ &  4\% \\
  Uncertainty in $D_1(2420)^0$ and $D^*_2(2460)^0$ fit &
  $(0.3 - 7.4)\%$   \\
 \hline
 \end{tabular}
 \end{center}
\end{table}

\section{SUMMARY}
\label{sec:Summary}

We present preliminary results from the study of
$\Bm$ decays to $D^{*+}\pim\pim$ and $D^{+}\pim\pim$.
The inclusive branching fractions for $\Bm\to\Dspipi$ and 
$\Bm\to\Dpipi$, and the exclusive branching fractions for 
$(\Bm\to D_1(2420)^0\pim) \times (D_1(2420)^0\to D^{*+}\pim)$,
$(\Bm\to D_2^*(2460)^0\pim) \times (D_2^*(2460)^0\to D^{*+}\pim)$,
and
$(\Bm\to D_2^*(2460)^0\pim) \times (D_2^*(2460)^0\to D^{+}\pim)$
are summarized in Table~\ref{tab:summary} and compared with 
results from BELLE and CLEO. Good agreement is seen between all
three experiments for all five results. Determinations of $R$ 
(see Eq.~\ref{eq:R})
are also in agreement among the three experiments and are 
consistent with the range expected in Ref.~\cite{leibovich}.
However, the measurements of $R$ differ significantly 
(by a factor of $\sim 2$) from the expectation of $R \approx 0.35$ given in 
Ref.~\cite{neubert}, and may thus provide some discrimination between 
the various HQET based calculations.

\begin{table}[!htbp]
  \begin{center}
    \caption{Measurements of the branching fractions for 
      \Bm decays to the
      $D^{*+}\pi^-\pi^-$ and $D^{+}\pi^-\pi^-$ final states.
      The first uncertainty is statistical, the second is systematic.
      The third uncertainty in the BELLE results is an additional
      systematic due to choice of selection criteria. Uncertainties
      on $R$ for BELLE and CLEO are the sum in quadrature of the 
      statistical and systematic uncertainties.
      \label{tab:summary}}
    \begin{tabular}{l|c|c|c}
      \hline
                 & \babar & BELLE~\cite{belle-prd} 
                 & CLEO~\cite{cleo-conf99-6} \\
                 & (preliminary) &               &   \\
      \hline
       & \multicolumn{3}{c}{Branching Fraction $(10^{-3})$}\\
      \hline
      $ \Bm \to D^{*+}\pim\pim$  & 
      {\boldmath $1.22 \pm 0.05 \pm 0.18$} & 
      $1.24 \pm 0.08 \pm 0.22$ & 
      $1.9 \pm 0.7 \pm 0.3$ \\
      $\Bm \to D^{+}\pim\pim$  & 
      {\boldmath $0.87 \pm 0.04 \pm 0.13$}&
      $1.02 \pm 0.04 \pm 0.15$ & 
      $< 1.4$ (90\% C.L.)\\
      \hline
      $(\Bm\to D_1(2420)^0\pim)$ &&&\\
      \hskip 3mm $\times\, (D_1(2420)^0\to D^{*+}\pim)$ & 
      {\boldmath $0.59 \pm 0.03 \pm 0.11$} & 
      $0.68 \pm 0.07 \pm 0.13 \pm 0.03$ & 
      $0.69^{+0.18}_{-0.14} \pm 0.12$ \\
      $(\Bm\to D_2^*(2460)^0\pim)$&&&\\
      \hskip 3mm $\times\,  (D_2^*(2460)^0\to D^{*+}\pim)$ & 
      {\boldmath $0.18\pm 0.03 \pm 0.05$}& 
      $0.18 \pm 0.03 \pm 0.03 \pm 0.02$&
      $0.31\pm 0.08 \pm 0.05$ \\
      \hline
      $(\Bm\to D_2^*(2460)^0\pim)$ &&&\\
      \hskip 3mm $\times\, (D_2^*(2460)^0\to D^{+}\pim)$ &
      {\boldmath $0.29\pm 0.02 \pm 0.05$} &
      $0.34\pm 0.03 \pm 0.06 \pm 0.04$&
      --- \\
      \hline
       & \multicolumn{3}{c}
       {$R \equiv \Br(\Bm \to D_2^*(2460)^0\pim)/\Br(\Bm 
           \to D_1(2420)^0\pim)$}\\
      \hline
      & {\boldmath $0.80 \pm 0.07 \pm 0.16$} & 
      $0.77 \pm 0.15$ & $1.8 \pm 0.8$ \\
      \hline
    \end{tabular}
  \end{center}
\end{table}

\section{ACKNOWLEDGMENTS}
\label{sec:Acknowledgments}

We are grateful for the 
extraordinary contributions of our \pep2\ colleagues in
achieving the excellent luminosity and machine conditions
that have made this work possible.
The success of this project also relies critically on the 
expertise and dedication of the computing organizations that 
support \babar.
The collaborating institutions wish to thank 
SLAC for its support and the kind hospitality extended to them. 
This work is supported by the
US Department of Energy
and National Science Foundation, the
Natural Sciences and Engineering Research Council (Canada),
Institute of High Energy Physics (China), the
Commissariat \`a l'Energie Atomique and
Institut National de Physique Nucl\'eaire et de Physique des Particules
(France), the
Bundesministerium f\"ur Bildung und Forschung and
Deutsche Forschungsgemeinschaft
(Germany), the
Istituto Nazionale di Fisica Nucleare (Italy),
the Foundation for Fundamental Research on Matter (The Netherlands),
the Research Council of Norway, the
Ministry of Science and Technology of the Russian Federation, and the
Particle Physics and Astronomy Research Council (United Kingdom). 
Individuals have received support from 
the A. P. Sloan Foundation, 
the Research Corporation,
and the Alexander von Humboldt Foundation.


\begin{thebibliography}{99}

\bibitem{hqshqet} N. Isgur and M. B. Wise, ``Heavy Quark Symmetry,'' in
       {\it $B$ Decays, Revised 2nd Edition}, edited by S. Stone (World
       Scientific, Singapore), p.231 (1994); 
       M. Neubert, Phys. Reports {\bf 245}, 259 (1994); 
       J. Richman and P. Burchat, Rev. Mod. Phys., {\bf 67}, 893 (1995).

\bibitem{barteltshukla} J. Bartelt and S. Shukla, ``Charmed Meson 
Spectroscopy'', Ann. Rev. Nucl. Part. Sci., 45:133 (1995).

\bibitem{infin_mass} A. De Rujula, H. Georgi, and S. L. Glashow,
        Phys. Rev. Lett. {\bf 37}, 785 (1976); 
        H. J. Schnitzer, Phys. Rev. D {\bf 18}, 3482 (1978); 
        D. Pignon and C. A. Picketty, Phys. Lett. B {\bf 81}, 334 (1979); 
        S. Godfrey and R. Kokoski, Phys. Rev. D {\bf 43}, 1679 (1991).

\bibitem{isgurwise89} N. Isgur and M. B. Wise, 
        Phys. Lett. B {\bf 232}, 113 (1989).

\bibitem{cleo-conf99-6}
  CLEO Collaboration, S.~Anderson {\it et al.}, 
   CONF 99-6 (1999), \\
  {\tt http://www.lns.cornell.edu/public/CONF/1999/CONF99-6/DJ\_lp99.ps}.


\bibitem{prevmeas} ARGUS Collaboration, H. Albrecht {\it et al.}, 
        Phys. Rev. Lett. {\bf 56}, 549 (1986); 
        ARGUS Collaboration, H. Albrecht {\it et al.}, 
        Phys. Lett. B {\bf 221}, 422 (1989); 
        ARGUS Collaboration, H. Albrecht {\it et al.}, 
        Phys. Lett. B {\bf 232}, 398 (1989); 
        Tagged Photon Spectrometer Collaboration (FNAL E691), 
        J. C. Anjos {\it et al.}, 
        Phys. Rev. Lett. {\bf 62}, 1717 (1989);         
        CLEO Collaboration, P. Avery {\it et al.}, 
        Phys. Rev. D {\bf 41}, 774 (1990); 
        E687 Collaboration, P. L. Frabetti {\it et al.}, 
        Phys. Rev. Lett. {\bf 72}, 324 (1994);  
        CLEO Collaboration, P. Avery {\it et al.}, 
        Phys. Lett. B {\bf 331}, 236 (1994) 
        [Erratum-ibid. B {\bf 342}, 453 (1995)]; 
        A. E. Asratyan {\it et al.}, 
        Z. Phys. C {\bf 68}, 43 (1995). 
        OPAL Collaboration, K. Ackerstaff {\it et al.}, 
        Z. Phys. C {\bf 76}, 425 (1997). 
        ALEPH Collaboration, D. Buskulic {\it et al.}, 
        Z. Phys. C {\bf 73}, 601 (1997). 
        DELPHI Collaboration, D. Bloch {\it et al.}, 
        Phys. Lett. B {\bf 426}, 231 (1998); 

\bibitem{PDG} Particle Data Group,
        K. Hagiwara {\it et al.}, 
        Phys. Rev. {\bf D 66}, 010001 (2002).

\bibitem{belle-conf-0235}
  BELLE Collaboration, K.~Abe {\it et al.}, BELLE-CONF-0235, 
  submitted to ICHEP02, Amsterdam (July 2002).

\bibitem{belle-prd} BELLE Collaboration, K.~Abe {\it et al.}, 
        submitted to Phys. Rev. D, hep-ex/0307021.

\bibitem{leibovich} A. K. Leibovich, Z. Ligeti, I. W. Stewart, M. B. Wise, 
        Phys. Rev. D {\bf 57}, 308 (1997).

\bibitem{neubert} M. Neubert, 
        Phys. Lett. B {\bf 418}, 173 (1998).

\bibitem{babarnim} \babar\ Collaboration, B.~Aubert {\it et al.}, Nucl.\
Instrum.\ and Methods A {\bf 479}, 1 (2002).

\bibitem{foxwolfram} G. C. Fox and S. Wolfram, 
        Phys. Rev. Lett. {\bf 41}, 1581 (1978).

\end{thebibliography}
\end{document}